\documentclass[final,5p,times,twocolumn]{elsarticle} 

\usepackage{amssymb}
\usepackage{amsmath}
\usepackage{siunitx}
\usepackage{subcaption}
\usepackage{cleveref}
\crefname{figure}{Figure}{Figures}

\usepackage{lineno}

\journal{Nuclear Physics A}

\begin{document}

\begin{frontmatter}

\title{Exploring the Design and Measurements of Next-Generation 4H-SiC LGADs} %% Article title

\author[FJFI,FZU]{Peter Švihra\corref{cor}} %% Author name
\ead{peter.svihra@cern.ch}
\ead[url]{https://cern.ch/peter-svihra}
\cortext[cor]{Corresponding author}

\author[onsemi]{Jan Chochol}
\author[FJFI]{Vladimír Kafka}
\author[onsemi]{Adam Klimsza}
\author[onsemi]{Adam Kozelsky}
\author[FZU]{Jiří Kroll}
\author[onsemi]{Roman Malousek}
\author[FJFI]{Mária Marčišovská}
\author[FJFI]{Michal Marčišovský}
\author[FZU]{Marcela Mikeštíková}
\author[CERN]{Michael Moll}
\author[onsemi]{David Novák}
\author[FJFI]{Radek Novotný}
\author[onsemi]{Peter Slovák}
\author[onsemi]{Radim Špetík}
\author[CERN]{Moritz Wiehe}

\affiliation[FJFI]{organization={Department of Physics, FNSPE CTU in Prague},
            addressline={Brehova 78/7},
            city={Prague},
            postcode={115 19},
            country={Czechia}}

\affiliation[FZU]{organization={Department of Detector Development and Data Processing, FZU CAS},
            addressline={Na Slovance 1999/2},
            city={Prague},
            postcode={182 00},
            country={Czechia}}

\affiliation[onsemi]{organization={onsemi},
            addressline={1. máje 2230},
            city={Rožnov pod Radhoštěm},
            postcode={756 61},
            country={Czechia}}
            
\affiliation[CERN]{organization={EP-DT-DD, CERN},
            addressline={Esplanade des Particules 1},
            city={Geneva},
            postcode={1211},
            country={Switzerland}}

%% Abstract
\begin{abstract}
This contribution presents the design, production, and initial testing of newly developed 4H-SiC Low Gain Avalanche Detectors (LGADs). The evaluation includes performance metrics such as the internal gain layer’s efficiency in enhancing signal generation. Initial laboratory and Transient Current Technique (TCT) measurements provide insight into the device’s stability and response to the signal.

Due to the increase of availability provided by the industry, 4H-SiC is emerging as a strong candidate for the next-generation of semiconductor detectors. Such sensors are promising due to the inherent radiation tolerance of 4H-SiC and its stable operation across a wide temperature range. However, due to the wider-bandgap of 4H-SiC compared to standard silicon, and difficulty to produce high-quality layers thicker than 50 \textmu m, an internal charge multiplication layer needs to be introduced.

The presented 4H-SiC LGADs, fabricated by onsemi, are optimized for an N-type substrate/epi wafer.The initial TCT and laboratory test results demonstrate fast charge collection and uniform multiplication across multiple samples produced on a single wafer.
\end{abstract}

\begin{graphicalabstract}

\section*{Introduction}
Re-emerging as a strong candidate for the next-generation semiconductor detectors, 4H-SiC promises an excellent performance due to its inherent advantages, including high radiation tolerance and the ability to operate across a broad temperature range without significant annealing effects.
However, the potential disadvantage for particle detection comes from its wider bandgap where signals generated in 4H-SiC detectors are inherently lower than those produced by traditional silicon detectors.
This is also more pronounced due to the difficulty of processing thick sensors which stably reach thicknesses only up to \SI{50}{\micro\meter}.

To address this, a charge multiplication layer can be implemented, compensating for the lower signal generation.
The 4H-SiC LGADs discussed here, fabricated by onsemi, are designed specifically on an N-type substrate/epi wafer, with the gain layer implanted approximately $1~\mathrm{\mu m}$ below the surface. These first-generation LGADs with dimensions of 3x3 mm$^2$ were produced in early 2024 and have since undergone extensive laboratory testing.

\begin{center}
\includegraphics[width=1.\linewidth]{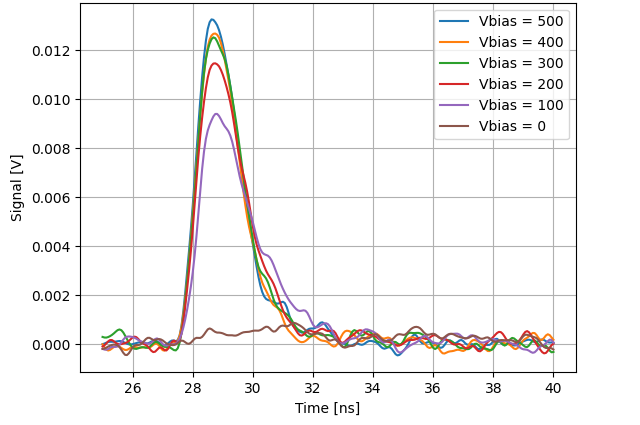}
\hfill
\includegraphics[width=1.\linewidth]{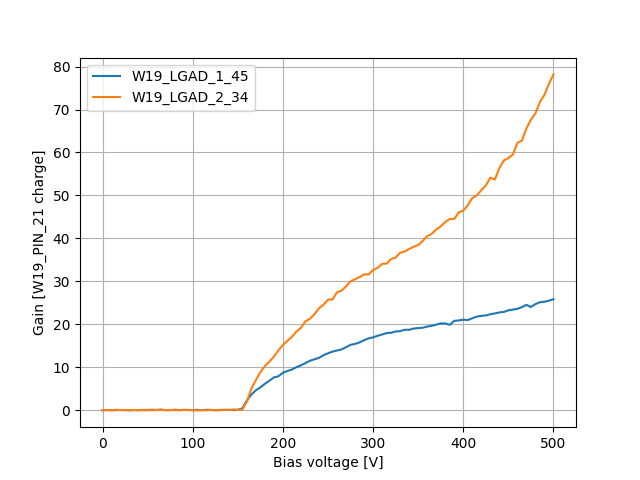}
Voltage dependencies of TCT response of 4H-SiC PN diode W19\_PN\_21 (left) and of calculated gain as a ratio of total signal measured by LGADs and PN (right).
\end{center}

\section*{Characterisation}
As the first step, wafers with different ranges of doping concentrations were produced and evaluated with IV and CV scans.
Each of the wafers contained three types of devices - two LGAD diodes with varied gain and reference diode with no gain.
Preliminary electrical performance results from one wafer indicate excellent uniformity.

% \caption{Signal response of a 4H-SiC diode with no-gain to UV pulse under different bias voltage. Each line is averaged from multiple measurements.}

Afterwards, initial TCT and beta-source measurements were performed which revealed fast charge collection times and provided details on charge multiplication across the active area.
The results align well with predictions from TCAD simulations.

\section*{Outlook}
Further testing of the devices is planned, along with the evaluation of an already performed proton irradiation campaign targeting proton fluences up to \SI{1e16}{1~\mega \electronvolt~n.eq}.

\end{graphicalabstract}

\begin{highlights}
\item First systematic electrical characterization of 4H-SiC LGADs

% This study presents the first systematic characterization of 4H-SiC Low Gain Avalanche Detectors (LGADs), fabricated using an ion-implanted internal gain layer. IV/CV measurements confirm stable operation at high voltages with low leakage currents, while Transient Current Technique (TCT) and beta-source tests verify charge multiplication and sub-100 ps timing response. These findings demonstrate the viability of 4H-SiC as a next-generation semiconductor material for radiation-tolerant, precision timing applications.

\item High production yield and stability of 4H-SiC LGADs from electrical testing

\item First particle detection using silicon carbide LGADs

\item Matching gain measurement in from TCT and beta source tests

\item Timing precision of silicon carbide LGADs down to 100 ps

% IV/CV measurements indicate that approximately \SI{85}{\percent} of tested 4H-SiC LGAD devices exhibit stable electrical performance, with low leakage currents and breakdown voltages exceeding 500 V. The consistency of full depletion voltages across multiple wafer samples further confirms the reproducibility of the fabrication process. These results suggest that 4H-SiC LGADs can achieve reliable large-scale production, a crucial step toward their integration into high-energy physics experiments and other radiation-intense applications.

\end{highlights}

%% Keywords
\begin{keyword}
4H-SiC \sep silicon carbide \sep LGAD \sep TCT \sep wide-bandgap semiconductor \sep ionizing radiation detector


\end{keyword}

\end{frontmatter}

% \linenumbers
\section{Introduction}
The development of next-generation radiation-tolerant semiconductor detectors is driven by the requirements of high-energy physics experiments and other applications operating in challenging environments.
Silicon-based Low Gain Avalanche Detectors (LGADs) have already demonstrated exceptional time resolution with moderate internal gain, making them highly suitable for tracking applications at collider experiments \cite{PELLEGRINI201412,SADROZINSKI201618}.
However, the increasing availability of high-quality 4H-SiC wafers -- driven in part by their adoption in the power electronics industry, which has helped reduce costs -- has generated considerable interest in extending LGAD technology to wide-bandgap semiconductor materials, enabling broader academic research.

Compared to silicon (with a bandgap of approximately \SI{1.12}{eV}), 4H-SiC possesses a significantly larger bandgap (approximately \SI{3.26}{eV}).
It also indicates superior radiation tolerance (current estimates indicate threshold displacement energy above \SI{24}{\electronvolt} \cite{Weimin}), higher breakdown voltages (critical field factor 10 larger than in silicon \cite{Konstantinov}), and stable performance across a broader temperature range.

Despite these inherent benefits, leveraging 4H-SiC for precision timing and tracking applications necessitates an internal gain mechanism due to its relatively low intrinsic carrier generation rates.
To achieve this, ion implantation techniques were utilized to define specialized doping profiles that form an internal multiplication region, converting a standard diode structure into an LGAD.
This implantation-based process closely mirrors standard silicon detector fabrication methods and differs from the epitaxial growth approach described in previous studies \cite{Zhao,Yang}.
The present work focuses on the characterization of recently fabricated 4H-SiC LGADs, emphasizing their electrical stability, Transient Current Technique (TCT) performance, and response to a beta radiation source.

\section{Design and Fabrication of 4H-SiC LGADs}
The devices under investigation were fabricated by onsemi utilizing 4H-SiC 6 inch wafers featuring N-type substrates with epitaxial layers of \SIlist{30;50}{\micro\meter} thickness, separated from the substrate using highly doped buffer layer \cite{novotny}.
A specifically engineered internal gain layer was introduced on the front side of each diode to facilitate controlled avalanche multiplication.
Multiple wafers were produced, each with varying doping concentrations in the gain layer to systematically evaluate device performance.

Each wafer featured a set of \numproduct{3 x 3}\si{\milli\meter\squared} devices, including distinct LGAD variants (LGAD1 and LGAD2), differing primarily in the doping concentration and thus the magnitude of internal gain, alongside the standard PN diodes without internal gain for comparative evaluation.
The top side of devices was either covered with full metallisation or a metal grill structure (alternating metal and no-metal lines) designed to enable efficient detection of UV light.
Edge termination of the devices was implemented through a Junction Termination Extension (JTE) scheme, optimized specifically for applications exceeding breakdown voltages of \SI{1}{\kilo\volt}, reflecting techniques typically employed in power electronics.

\section{Experimental Setup}
This contribution focuses on evaluation of performance of samples from a single wafer W19.
However, overall plots of measured breakdowns are provided across multiple measured wafers with epitaxial thickness \SI{30}{\micro\meter} (W16) and  \SI{50}{\micro\meter} (W17 and W19).
Labelling of the individual samples was done as "W<wafer>\_<type>\_<index>".

\subsection{IV and CV Measurements}
Electrical characterization was carried out using a manual probe-station equipped with a source measure unit capable of safely reaching up to \SI{500}{\volt} without dry air.
While as source measure unit we have used Keithley 2657A, the probe station was equipped only with a standard coaxial-mounted chuck, biasing via needle and lacking proper shielding.
Current-voltage (IV) measurements were performed to assess the leakage current and breakdown voltage while capacitance-voltage (CV) measurements helped quantify the depletion voltage and verify the consistency of doping profiles among different sensors.
All measurements were performed under reverse bias conditions.

\subsection{Transient Current Technique (TCT)}
TCT measurements were conducted by illuminating the top surface of the device with laser pulses of \SI{375}{\nano\meter} and sub-nanosecond duration.
Fast readout electronics (\SIrange{0.1}{2}{\giga\hertz}, \SI{44}{\decibel} Cividec current amplifier and \SI{2.5}{\giga\hertz} oscilloscope) were used to capture the transient signals, from which the charge and drift profiles were reconstructed.

\subsection{Beta-Source Setup}
A $^{90}\mathrm{Sr}$ source emitting $\beta$-particles was placed above a stack of two previously tested silicon LGADs and 4H-SiC device under test.
The readout chain included three low-noise current-sensitive amplifier and an oscilloscope \cite{CurrasRivera:2291517}.
Waveforms were recorded for each particle detected in coincidence across all three devices, and offline analysis was used to extract key parameters such as pulse shape and duration.
The timing resolution can then be obtained for all three measured sensors using the approach described in \cite{mckarris_2019_tshk4-3nj72}.

\begin{figure}[p]
    \centering
    \begin{minipage}{.41\textwidth}
        \centering
        \includegraphics[width=1.\linewidth]{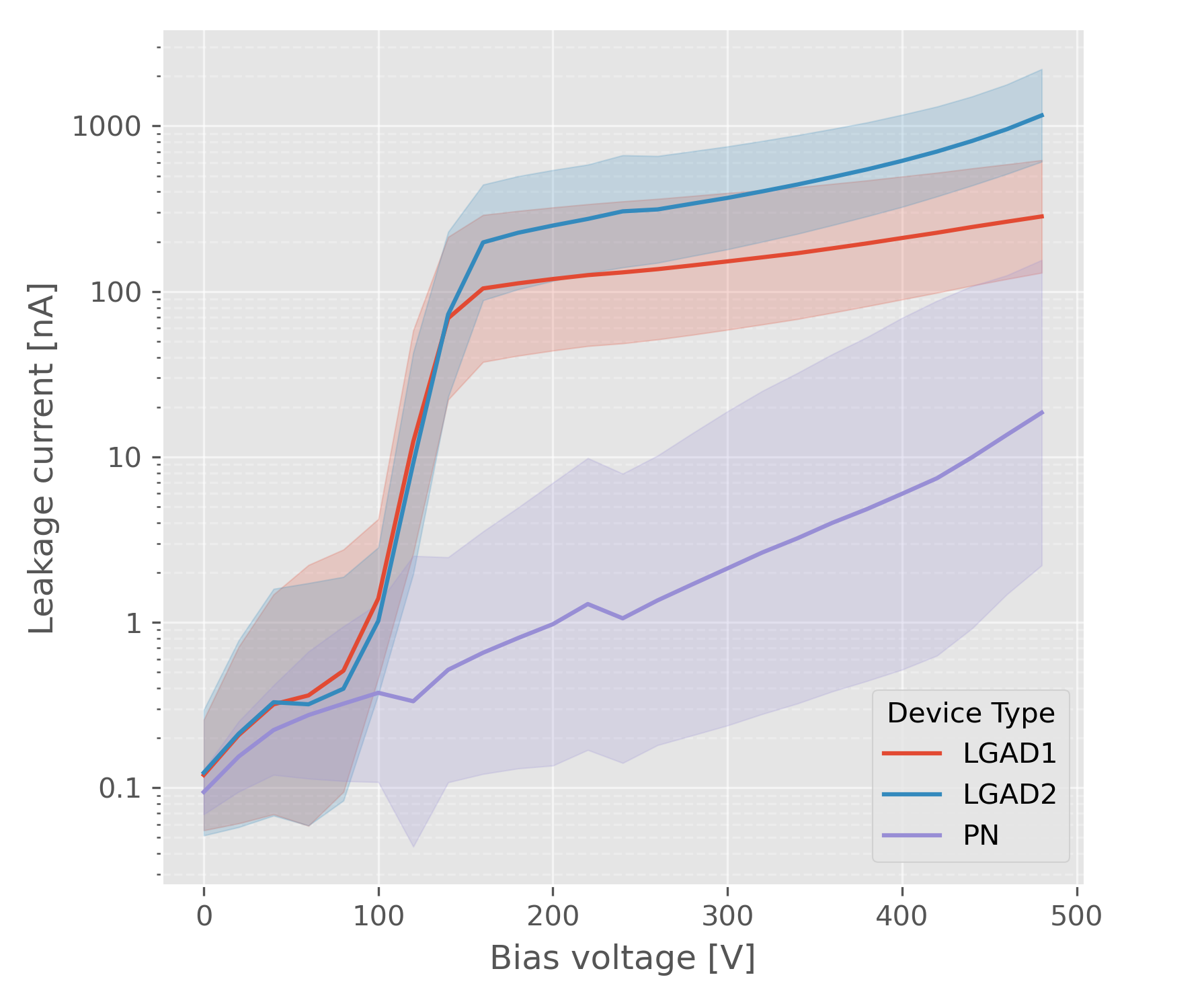}
        \caption{Comparison of current on voltage dependencies for no-gain PN diode and two types of LGADs from W19. Each type measured for around 20 samples plotted as mean (solid) with stdev (shaded). Sharp increase above \SI{100}{\volt} matches gain layer depletion.}
        \label{fig:IV}
    \end{minipage}
    \hfill
    \begin{minipage}{.43\textwidth}
        \centering
        \includegraphics[width=1.\linewidth]{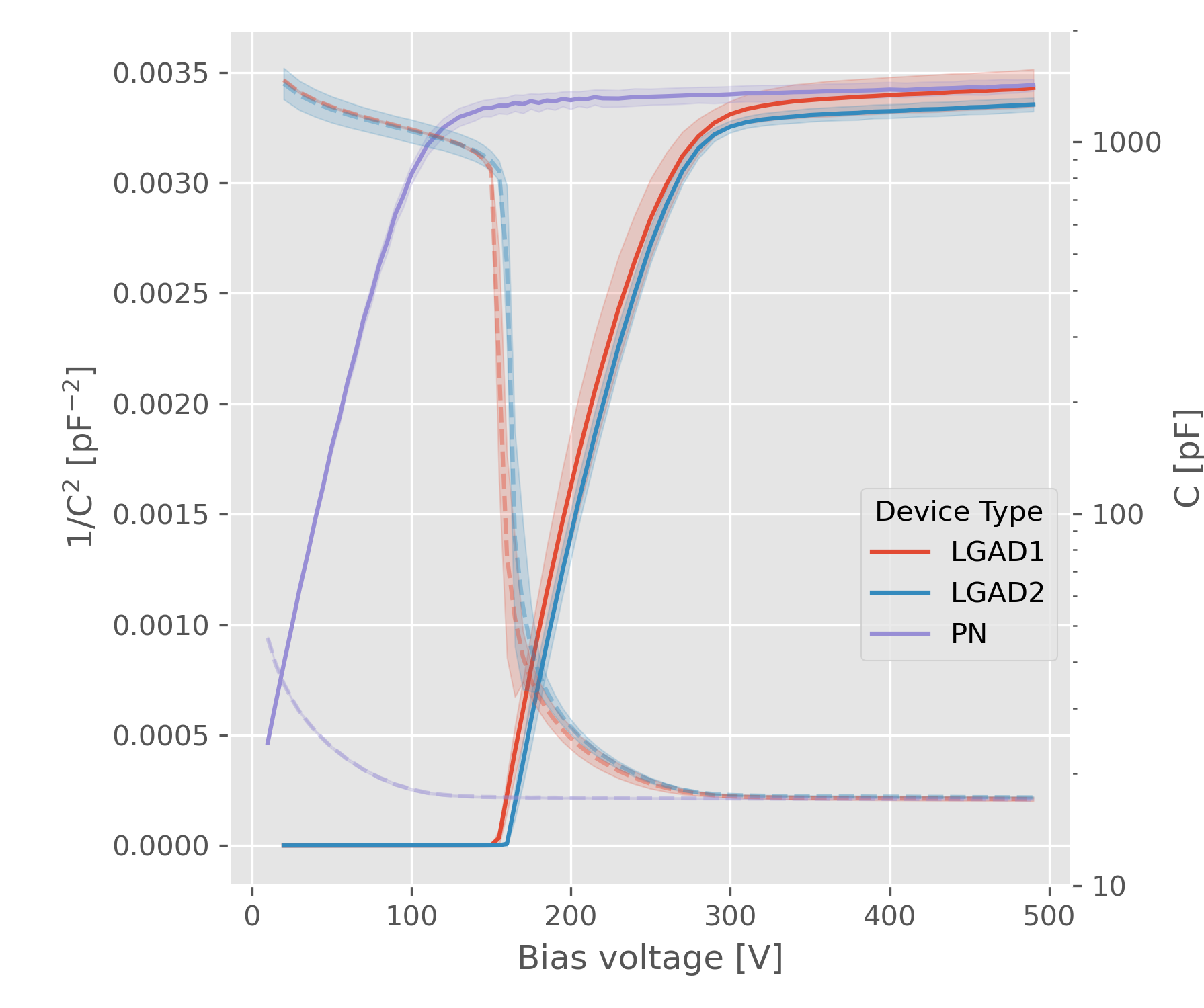}
        \caption{Comparison of voltage dependency of $1/C^2$ (left axis, solid) and $C$ (right axis, dashed) for no-gain diode and two types of LGADs from W19. Each type measured for around 20 samples plotted as mean with stdev (shaded).}
        \label{fig:CV}
    \end{minipage}
    \begin{minipage}{.45\textwidth}
        \centering
        \includegraphics[width=1.\linewidth]{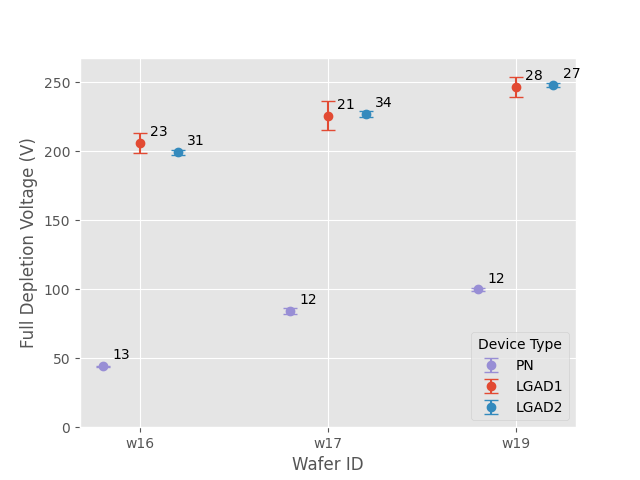}
        \caption{Full depletion voltage for different device types (PN, LGAD1, LGAD2) across three wafers (\SI{30}{\micro\meter} thick W16, and \SI{50}{\micro\meter} thick W17 and W19). Each point represents the fitted full depletion voltage with error bars indicating standard deviation and adjacent values specifying the number of measured samples.}
        \label{fig:depletion}
    \end{minipage}
\end{figure}

\section{Results}
\subsection{IV and CV Measurements}
Approximately \SI{85}{\percent} of all tested devices demonstrated reliable performance in both IV and CV measurements.
Specifically, for W19, the good quality devices exhibited reverse leakage currents below the microampere level at typical bias voltages between \SIlist{100;300}{\volt}, see \cref{fig:IV}.
Although the measured currents slightly exceed typical SiC limits, this is due to limitations of the measurement setup rather than the material itself.
Despite this, the trends and comparison between PN and LGAD structures demonstrate excellent performance.
Furthermore, the breakdown voltage for most diodes was found to exceed \SI{500}{\volt}, indicating robust edge termination and a consistent fabrication process.

Measured CV characteristics revealed a stable depletion region extended across the entire active thickness at bias voltages in the \SIrange{200}{250}{\volt} range for LGADs and around \SIrange{50}{100}{\volt} for PN diodes, see \cref{fig:CV}, consistent with design expectations.
The expected and observed difference in full depletion voltage between LGADs and standard PN diodes arises because the thin gain layer in LGADs has to deplete first, at around \SI{150}{\volt}, before the rest of the epitaxial layer becomes fully depleted.
The gradual kink before full depletion is most probably caused by highly doped buffer layer that separates substrates from EPI.

By fitting the turn-on and stable slopes for each devices $1/C^2$ values, full-depletion voltages were obtained and are plotted in \cref{fig:depletion} for three wafers and all device types.

\subsection{TCT Analysis}
\begin{figure}[p]
\begin{minipage}{.47\textwidth}
    \begin{subfigure}[t]{\linewidth}
        \centering
        \includegraphics[width=1.\linewidth]{figures/tct_PIN_signal.png}
        \caption{W19\_PN\_21 (no gain).}
    \end{subfigure}
    \begin{subfigure}[t]{\linewidth}
        \centering
        \includegraphics[width=1.\linewidth]{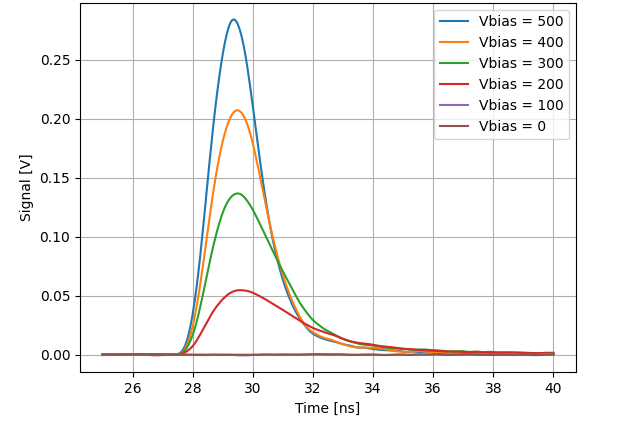}
        \caption{W19\_LGAD1\_45 (gain).}
    \end{subfigure}
    \caption{Transient signal response of a 4H-SiC diodes to UV pulse under different bias voltage. Each line is averaged from multiple measurements.}
    \label{fig:TCT}
    \end{minipage}
    \begin{minipage}{.47\textwidth}
        \centering
        \includegraphics[width=1.\linewidth]{figures/tct_gain.png}
        \caption{Voltage dependency of gain calculated as a ratio of total signal of selected W19 LGAD samples and no-gain W19\_PN\_21 diode from the same wafer. Measured using TCT, total signal calculated as integral of the transient pulse.}
        \label{fig:tct_gain}
    \end{minipage}
\end{figure}

For all subsequent tests single device of each type from W19 was selected and wire-bonded to a PCB for simpler handling.
These include no gain diode W19\_PN\_21, and two LGADs W19\_LGAD1\_45, and W19\_LGAD2\_34.
The TCT measurements confirmed that the use of the internal gain layer enhances the magnitude of both the transient current as well as total collected charge calculated as its integral (see \cref{fig:TCT}).
Resulting gain as function of voltage is presented in \cref{fig:tct_gain}, obtained as the ratio of the measured charge between PN and LGAD diodes.

\begin{figure}[t]
    \centering
    \includegraphics[width=.47\textwidth]{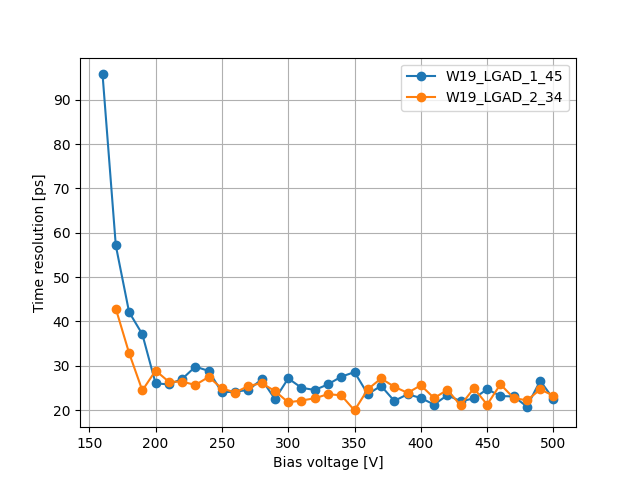}
    \caption{Voltage dependency of time resolution of W19 LGAD samples. Measured using TCT. This measurement does not reflect realistic timing performance -- charge injection is much higher than for a MIP, there are no Landau fluctuations, and laser pulse was used as time reference.}
    \label{fig:tct_timing}
\end{figure}

Additionally, the shape of the transient signals indicated a rapid charge collection time (see \cref{fig:tct_timing}), on the order of tens of picoseconds, uncalibrated to single minimum ionising particle (MIP) deposition, which is crucial for high-precision timing applications.

\subsection{Beta-Source Performance}
Using the same devices as for TCT, a preliminary analysis of signals generated by $\beta$-particles confirms that the internal multiplication layer enhances signal amplitude as expected, thereby improving the signal-to-noise ratio (SNR) compared to standard 4H-SiC no-gain PN diodes.

\begin{figure}[p]
    \centering
    \begin{minipage}{.47\textwidth}
        \centering
        \includegraphics[width=1.\linewidth]{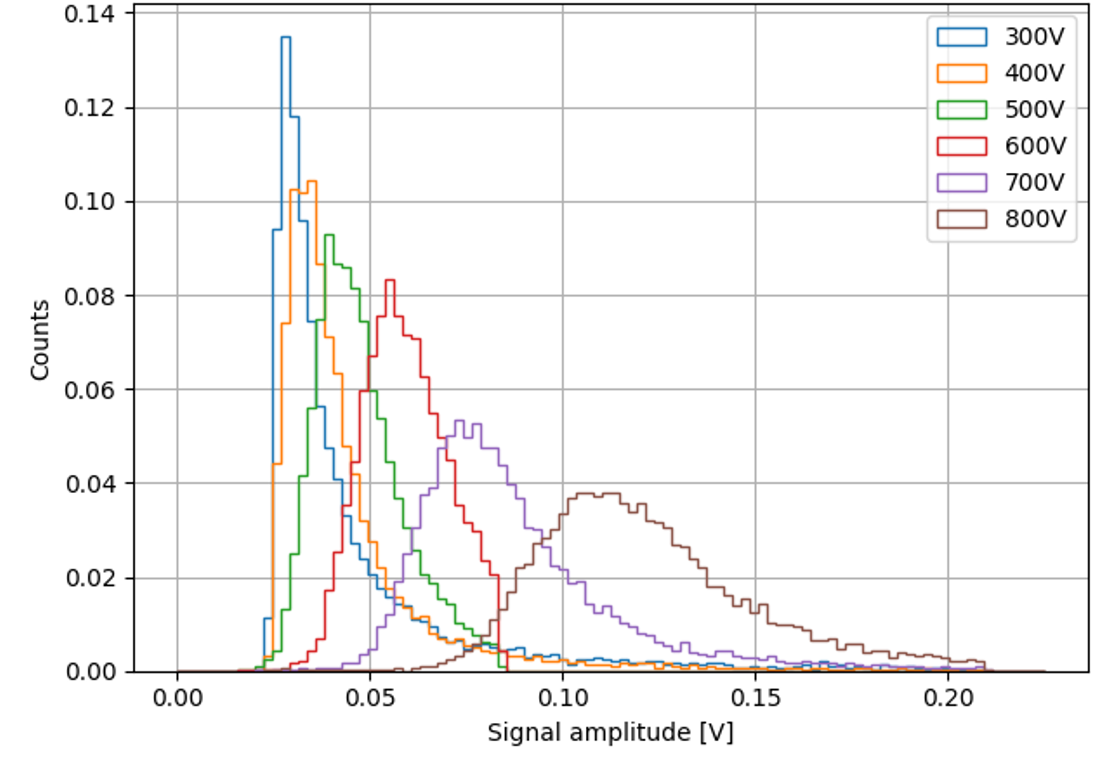}
        \caption{Distribution of 4H-SiC W19\_LGAD1\_45 signal response to $\beta$ particles under different bias voltage.}
        \label{fig:beta}
    \end{minipage}
    \begin{minipage}{.47\textwidth}
        \centering
        \includegraphics[width=1.\linewidth]{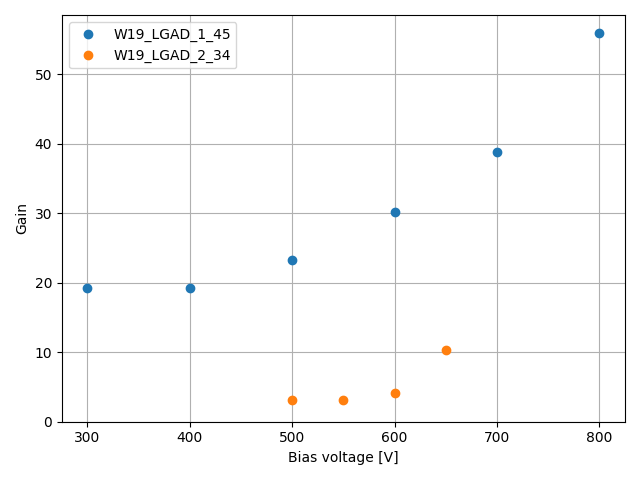}
        \caption{Voltage dependency of gain calculated as a ratio of total signal of W19 LGAD samples and theoretical prediction of MIP deposition of \SI{2550}{e^{-}} in \SI{50}{\micro\meter} 4H-SiC. Measured using beta-source.}
        \label{fig:beta_gain}
    \end{minipage}
    \begin{minipage}{.52\textwidth}
        \centering
        \includegraphics[width=1.\linewidth]{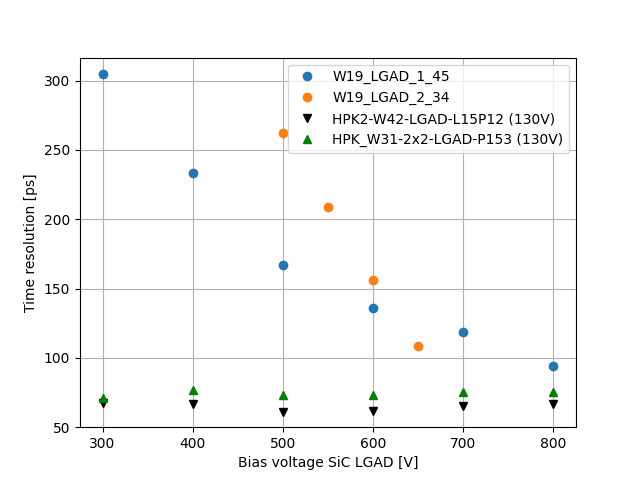}
        \caption{Voltage dependency of time resolution of W19 LGAD samples. Measured using beta-source, HPK LGADs shown for stability of the evaluation technique.}
        \label{fig:beta_timing}
    \end{minipage}
\end{figure}

The charge collected by the LGADs was estimated using the most probable value (MPV) obtained from a fit to the charge distribution shown in \cref{fig:beta}, employing a Landau-Gaussian convolution.
Integrated values were then converted to charge by dividing the number with oscilloscope input resistance \SI{50}{\ohm} and amplifier gain 158.5.
This value was compared to the expected MIP deposition of \SI{2550}{e^{-}} in \SI{50}{\micro\meter} 4H-SiC (estimated based on 51 e-h pairs produced per micrometer of MIP in 4H-SiC \cite{Bruzzi}), the resulting ratio plotted versus bias voltage in \cref{fig:beta_gain}.
While the gain values for W19\_LGAD1\_45 are consistent with those observed in TCT, W19\_LGAD2\_34 did not achieve the previously recorded performance, requiring further investigation.

The timing resolution of W19\_LGAD1\_45, shown in \cref{fig:beta_timing}, exhibits promising trends, reaching the sub \SI{100}{\pico\second} range when biased at \SI{800}{\volt}.

\section{Conclusions}
This study successfully demonstrated the fabrication and characterization of novel 4H-SiC LGADs, establishing a crucial step toward extending LGAD technology beyond silicon-based detectors.
Measurements of electrical characteristics confirmed stable operation at high voltages with low leakage currents, while TCT and beta-source evaluations verified the presence of internal charge multiplication and promising timing characteristics.
Notably, W19\_LGAD1\_45 achieved sub \SI{100}{\pico\second} timing performance and a gain around 20 consistent between TCT and beta-source measurements, whereas W19\_LGAD2\_34 exhibited lower-than-expected performance.
This might be due to different gain suppression mechanisms and requires further evaluation of samples.

The results highlight the potential of 4H-SiC LGADs as viable candidates for applications requiring high radiation tolerance, precise timing, and extended operational temperature ranges.
Future efforts will focus on optimizing doping concentrations, refining device geometries, and assessing long-term stability under extreme radiation conditions. Additionally, further test beam campaigns and Monte Carlo simulations will be necessary to quantify the impact of these sensors in experimental environments such as high-luminosity collider detectors.

\section*{Acknowledgements}
Researcher Peter Švihra conducts his research under the Marie Skłodowska-Curie Actions – COFUND project, which is co-funded by the European Union (Physics for Future – Grant Agreement No. 101081515).
This work was supported by the Technological Agency of the Czech Republic - Project TK05020011.
The team from the Institute of Physics of the Czech Academy of Sciences was supported via the projects LM2023040 CERN-CZ and FORTE - CZ.02.01.01/00/22\_008/0004632.

\section*{Declaration of generative AI and AI-assisted technologies in the writing process}
During the preparation of this work the author(s) used chatGPT in order to improve the readability of the paper. After using this tool/service, the author(s) reviewed and edited the content as needed and take(s) full responsibility for the content of the published article.

\bibliographystyle{elsarticle-num} 
\bibliography{cas-refs}

\end{document}